\documentclass[]{IEEEphot}

\usepackage{cite}
\usepackage[numbers,sort&compress]{natbib}

\begin{document}

\title{1.2 GHz Balanced Homodyne Detector for Continuous-Variable Quantum Information Technology}

\author{Xiaoxiong Zhang$^1$, Yichen Zhang$^1$, Zhengyu Li$^2$, Song Yu$^1$, Hong Guo$^2$}
\affil{$^1$State Key Laboratory of Information Photonics and Optical Communications, Beijing University of Posts and Telecommunications, Beijing 100876, China\\$^2$State Key Laboratory of Advanced Optical Communications Systems and Networks, School of Electronics Engineering and Computer Science, Center for Quantum Information Technology, Center for Computational Science and Engineering, Peking University, Beijing 100876, China}


\maketitle


\begin{receivedinfo}%
This work was supported by the Key Program of National Natural Science Foundation of China under Grant 61531003, the National Natural Science Foundation under Grant 61427813, the National Basic Research Program of China (973 Program) under Grant 2014CB340102, the China Postdoctoral Science Foundation under Grant 2018M630116, and the Fund of State Key Laboratory of Information Photonics and Optical Communications. Corresponding author: Yichen Zhang (e-mail: zhangyc@bupt.edu.cn); Song Yu (e-mail: yusong@bupt.edu.cn).
\end{receivedinfo}

\begin{abstract}%
Balanced homodyne detector (BHD) that can measure the field quadratures of coherent states has been widely used in a range of quantum information technologies. Generally, the BHD tends to suffer from narrow bands and an expanding bandwidth behavior usually traps into a compromise with the gain, electronic noise, and quantum to classical noise ratio, etc. In this paper, we design and construct a wideband BHD based on radio frequency and integrated circuit technology. Our BHD shows bandwidth behavior up to 1.2 GHz and its quantum to classical noise ratio is around 18 dB. Simultaneously, the BHD has a linear performance with a gain of 4.86k and its common mode rejection ratio has also been tested as 57.9 dB. With this BHD, the secret key rate of continuous-variable quantum key distribution system has a potential to achieve 66.55 Mbps and 2.87 Mbps respectively at the transmission distance of 10 km and 45 km. Besides, with this BHD, the generation rate of quantum random number generator could reach up to 6.53Gbps.

\end{abstract}


\section{Introduction}
The rapid development of quantum information technology requires effective and reliable methods to characterize the optical quantum states. In the application of nonclassical light, the balanced homodyne detector (BHD) has been found as an invaluable tool in measuring field quadratures of an electromagnetic mode \cite{kumar2012versatile}, and plays a significant role in quantum state detection with continuous variable \cite{lvovsky2009continuous,leonhardt1997measuring}. BHD, proposed by Yuen and Chan in 1983 \cite{yuen1983noise}, is now widely used in optical homodyne tomography \cite{kumar2012versatile}, establishment of Einstein-Podolsky-Rosen-type quantum correlations \cite{ou1992realization}, squeezed states detection \cite{grote2016high} as well as coherent states detection \cite{zhang2017continuous}. In recent years, BHD is employed to play a major role in continuous-variable quantum key distribution (CV-QKD) \cite{scarani2009security,gisin2002quantum} and vacuum-state-based quantum random number generator (QRNG) \cite{ma2016quantum,herrero2017quantum}.

CV-QKD has attracted increasing attention in the past few years, mainly as it uses standard telecom components, such as BHD, and don't need single photon detector. Various experiments have been undertaken, from laboratory environment reaching 80 km transmission distance with 1 MHz repetition rate in 2013 \cite{jouguet2013experimental}, to long distance field test, which has achieved a secret key rate of 7.3 kbps in the finite-size regime over a 50 km commercial fiber with a repetition rate of 5 MHz in 2017 \cite{zhang2017continuous}, showing a step forward in the development of CV-QKD. Comparing with discrete-variable quantum key distribution (DV-QKD), CV-QKD has advantages in higher secret key rate per pulse while suffers from low repetition rate \cite{chi2009high}. The repetition rate is limited mainly by the bandwidth of the BHD, high-speed data acquisition, and classical reconciliation scheme, in which BHD is the main limitation \cite{chi2009high}. On the other scene, a QRNG based on measuring the vacuum fluctuations has the advantages of high bandwidth, simple optical setup, insensitive to detection efficiency and multiple bits per sample \cite{zheng2018}. However, the final random number generation rate in this scheme is limited by the bandwidth of the BHD \cite{shen2010practical}. Therefore, as stepping towards commercial applications of CV-QKD and QRNG, there is a significantly growing demand in improving the bandwidth of BHD.

The design of BHD is normally based on three main criteria: a) the quantum to classical noise ratio (QCNR) should not be less than 10 dB \cite{QianqianZhou2010}; b) the common mode rejection ratio (CMRR) should be more than 30 dB \cite{QianqianZhou2010}; c) high bandwidth with a flat amplification gain \cite{kumar2012versatile}. Usually, the most tricky one is bandwidth, since it is affected by many factors, i.e., the terminal capacitance of the photodiode, the high-frequency performance of the trans-impedance amplifier (TIA) and the PCB layout. Considering the maturity of commercial photodiode and PCB layout technology, TIA has drawn our most attention. Referring to the most reported BHDs, the mentioned excellent operational amplifiers such as OPA847 (Texas Instruments) \cite{chi2011balanced,kumar2012versatile,chi2009high,zhou2015low,qin2016design,cooper2013high}, AD8015 (Analog Devices), LTC6409 (Linear Technology) \cite{duan2013300} and LTC6268 (Linear Technology) are more suitable for low-frequency applications according to the official datasheets. Thus with these chips, the QCNR and CMRR can be well optimized only in a narrow band, which will deteriorate when the bandwidth is expended. We are committed to finding a unique combination of the bandwidth, CMRR, and QCNR.

We operate a pair of near-identical high quantum efficiency photodiode and a resistor followed by a radio frequency (RF) voltage amplifier to build our BHD, whose bandwidth is 1.2 GHz which is superior to other excellent BHDs. What's more, the QCNR and CMRR have reached to 18.5 dB and 57.9 dB respectively so that the BHD almost meets the criteria mentioned above. With such a BHD, the CV-QKD system has potential to reach a higher secret key rate, and the random number generation rate of the QRNG will also be improved.

This paper is organized as follow: In section 2, we report the construction and performance of the BHD, including the schematic of the circuit and the diagram of the optical experiment, as well as the test result. In section 3, we show the applications of the BHD. we exhibit the simulation between the secret key rate and the transmission distance in CV-QKD and analyze the random number generation rate in QRNG. Our conclusions are drawn in section 4.

\section{Construction and performance}

In this section, We present the construction of our BHD based on RF technology and integrated circuit design. Besides, we also set up an optical experiment to test the performance of our BHD in the telecommunication wavelength region.

\subsection{Schematic representation}

As shown in Fig. 1 (a), we construct the circuit schematic including two reverse biased InGaAs photodiodes (PD) from Hamamatsu (G9801-32, bandwidth: 2 GHz, terminal capacitance: 1 pF, sensitivity ratio: 90\% typically and quantum efficiency: 72.2\% at 1550 nm) and a resistor followed by radio frequency integrated circuit (RFIC) from Agilent (ABA-52563, bandwidth: 3.5 GHz) and the values of components are also shown in the circuit schematic. One can build their own BHD based on the circuit schematic and the datasheet of the RFIC. Others schemes that use general TIA, such as OPA847 or LTC6409, have a perfect performance only at low frequency, typically tens and hundreds of MHz. As the frequency increases, there will be inflection points of critical parameters including the load capacitance, the common mode rejection ratio, the gain, and the output impedance, etc., that will result in deterioration of the TIA performance. Compared with general TIA schemes, the bandwidth of the BHD in our solution is determined by the response of PDs, the bandwidth of RFIC, and the cut-off frequency of the terminal capacitance along with the resistance. Whereas the PDs and the RFIC are never the primary limitations for the reason that a 50 $\Omega$ resistor and the parallel terminal capacitors of two PDs bring the upper bound of operating frequency to 1.59 GHz. If we consider the effect of the parasitic capacitance from the component package and the solder pins, the operating frequency will drop, and a total parasitic capacitance of 0.5 pF will result in a high-frequency loss of 0.3 GHz. To minimize the parasitic capacitance, we have taken some measures, including reducing the distance of the PD pins, shortening the length of traces on the PCB and using smaller package components. Eventually, the whole PCB is fixed in a metal box with electromagnetic shielding function.

The optical experiment setup is drawn in Fig. 1 (b). A 1550-nm fiber-coupled laser (NKT Basic E15, linewidth 100 Hz) offers continuous wave (CW) and the following variable optical attenuator (VOA$_1$) is used to adjust the beam power to an appropriate value. We operate an amplitude modulation (AM) from Photline (MXER-LN-10 with 10 GHz in 1550 nm) to modulate information generated by arbitrary waveform generator (AWG) to local oscillation (LO). Experimentally, one of the input ports of beam splitter (BS) is left unconnected to provide the vacuum state. The LO and the vacuum state will interfere at the BS with splitting ratio of 50:50. After that, there will be a VOA$_2$ and a variable optical delay (VOD) to balance the output arms of the fiber coupler so that a BHD only amplifies the differential signal. At the output of the BHD, we use spectrum analyzer (Rigol DSA815), oscilloscope (Keysight Technology MSOS804A) and FPGA with ADC (ADS5400, sampling rate: 1 GHz, sampling accuracy: 12 bits) to analyze the output signal in both frequency and time domain. This experimental system needs to be properly adjusted to test the following parameters of BHD.

\begin{figure}[t]
\centering
\includegraphics[width=35pc]{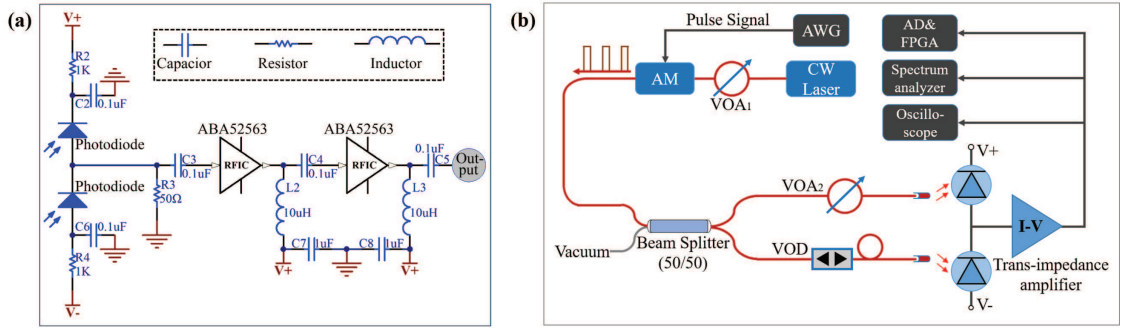}
\caption{(a) Simplified electronic circuit schematic of the balanced homodyne detector. The "V+" symbol represents the positive supply voltage while the "V-" symbol represents the negative supply voltage. The RFIC represents radio frequency integrated circuit whose model is ABA52563 from Agilent.  (b) Diagram of the optical experiment setup. The red lines are optical paths and the black lines are electrical cables. CW Laser: 1550 nm continuous wave fiber laser, AWG: Arbitrary waveform generator, VOA$_{1,2}$: Variable optical attenuators, VOD: Variable optical delay, AM: Amplitude modulator.}
\label{fig_env5}\vspace*{-6pt}
\end{figure}

\subsection{QCNR and bandwidth}
Quantum noise estimation is one of the most important processes in CV-QKD for the relevant physical quantities need to be calibrated in quantum noise units. And in QRNG, the quantum noise, from which the quantum random number generated is supposed to dominate the total noise in order to produce more bits per sample. Usually, the quantum noise level is required to be more than 10 dB above the classical noise level \cite{gray1998photodetector}. In order to point out that our BHD is fully applicable to CV-QKD and QRNG system, we have measured QCNR in the frequency and time domain with CW LO, from which we can also get the bandwidth information.
\subsubsection{Noise and bandwidth}
The optical experiment is demonstrated in Fig. 1 (b) except AM. In this case, the residual signal caused by the different quantum efficiency of PDs and the splitting ratio error of BS can be eliminated by adjusting VOA$_2$. In the case of the smallest residual signal, we record the electronic noise that is shown in Fig. 2 (a) in the time domain using oscilloscope and the root mean square value of electronic noise is 3.08 mV. While in the frequency domain, the background noise spectrum of the spectrum analyzer, the electronic noise spectrum of BHD, and the output noise spectrum of BHD under different LO power is shown in Fig. 2 (b). With the increase of LO power, the output noise power of the BHD will rise from kHz to 1.2 GHz and then drop sharply. In the low-frequency region, the lower cut-off frequency is determined by the DC blocking capacitor. However, in this region, the superimposed 1/f noise and instrument noise is so strong that the output noise spectrum of BHD is covered. As the frequency increases, the spectrum becomes unflattened ranging from 400 MHz to 600 MHz and 900 MHz to 1200 MHz due to signal integrity issues \cite{xu2012ultrastable}, which addresses high requirement on the PCB layout. An upper cut-off frequency appears at near 1.2 GHz and then the noise power rolls down until it drops to an electronic noise power level at 1.4 GHz nearby. It illustrates the validity of our methods to construct a $\sim$GHz high-speed BHD.

\begin{figure}[t]
\centering
\includegraphics[width=36pc]{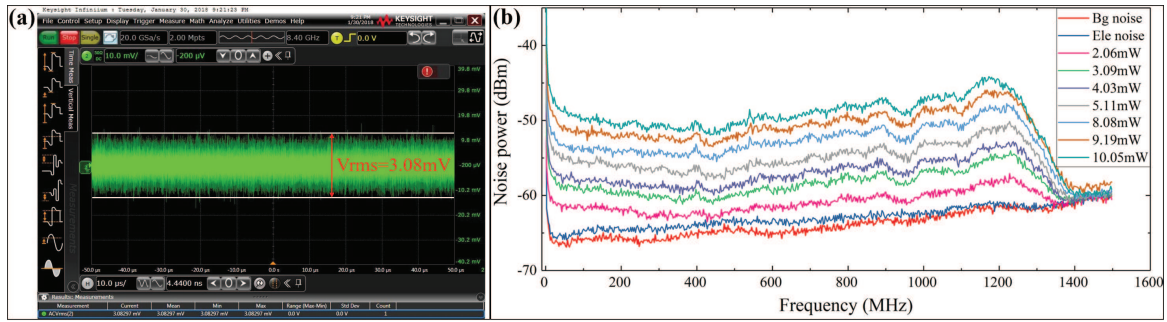}
\caption{(a) Root mean square (rms) electronic noise of the BHD. The Vrms value is 3.08 mV. (b) Measured noise power of BHD ranges from kHz to 1.5 GHz. Spectrum analyzer background noise spectrum (Bg noise curve), BHD electronic noise spectrum (Ele noise curve) and BHD noise spectrum at CW LO powers of 2.06, 3.09, 4.03, 5.11, 8.08, 9.19 and 10.05 mW (from the third lowest to highest curve). Resolution bandwidth: 100 kHz.}
\label{fig_env5}\vspace*{-6pt}
\end{figure}

\subsubsection{QCNR in frequency and time domain}
In the process of QCNR estimation, we prefer CW LO as an optical source. Since the electronic noise and quantum noise follow Gaussian distribution \cite{haw2015maximization} and the BHD is AC-coupled, the mean of the noise is zero and the variance of noise is equal to the noise power. We are required to verify the behavior of QCNR between frequency and time domain. Note that, in the following statements, the total noise is contributed by electronic noise and quantum noise. The electronic noise, that includes background noise of instrument and electronic noise of BHD is dominated by electronic noise of BHD. The quantum noise is calculated by subtracting the electronic noise from total noise.

We have got QCNR information in Fig. 2 (b), where one can only read the QCNR at a single frequency while the BHD is used to detect the quadrature of coherent state in the time domain. In this case, the spectral information needs to be transformed into a more readable form. We perform calculation by integrating the area under each curve in Fig. 2 (b) from 5 MHz to 1.2 GHz, since low-frequency noise (might be contributed by 1/f noise and instrument noise) is not mainly contributed by electronic noise and quantum noise \cite{chi2009high}. Therefore we get electronic noise power and a set of total noise power related to LO power. As indicated in Fig. 3 (a), it is evident that the quantum noise power has an approximately linear relationship with LO power and the QCNR is 18.5 dB at an LO power of 8.08 mW.

Similarly, the optical experiment is repeated in the time domain. With an oscilloscope, we set the sampling rate to 10 GSa/s, and set time base to 1 us to ensure that enough data is stored. With these test conditions, we measure the electronic noise voltage when LO is blocked and the total output noise voltage under different LO power. During data processing, we select one number per 10 numbers in the raw data to avoid correlation. Then we calculate the electronic noise variance and total noise variance. The quantum noise variance is also calculated by subtracting the electronic noise variance from total noise variance. As indicated in Fig. 3 (b), the quantum noise variance has an approximately linear relationship with LO power and a QCNR is up to 17.8 dB at an LO power of 8.05 mW, which has a good agreement with that in the frequency domain.

\begin{figure}[t]
\centering
\includegraphics[width=35pc]{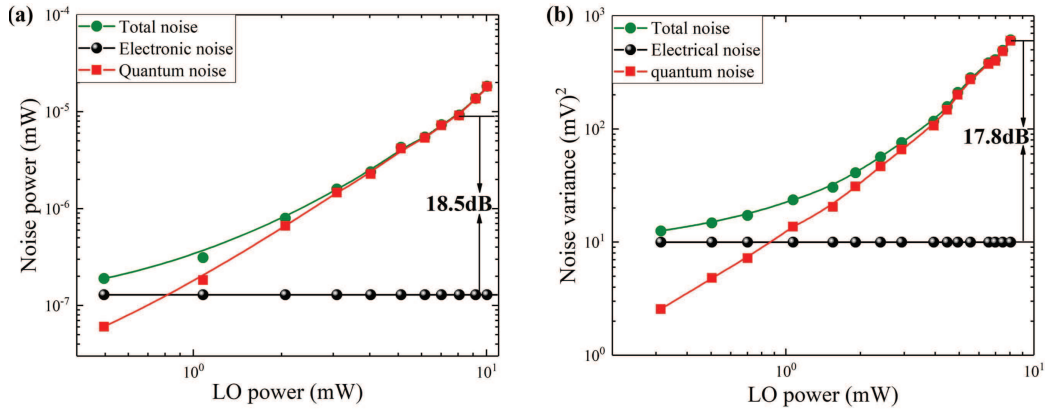}
\caption{(a) BHD noise power as a function of CW LO in the frequency domain. The quantum noise power to electronic noise power ratio is 18.5 dB at an LO power of 8.08 mW. (b) Noise variance as a function of the CW LO power in the time domain. The quantum noise variance to electronic noise variance ratio is 17.8 dB at an LO power of 8.05 mW. For both of the Figure (a) and (b), the green curve is total noise contributed by electronic noise and quantum noise. The black line is electronic noise from the BHD and the instrument. The red curve is quantum noise calculated by subtracting the electronic noise from the total noise.}
\label{fig_env5}\vspace*{-6pt}
\end{figure}

\subsection{Linearity and gain}

In CV-QKD, the true random numbers generated by QRNG are encoded to signal pulse at Alice's side and recovered by the BHD on Bob's side \cite{chi2011balanced}. As the output voltage should be proportional to the quadrature of the signal pulse, the BHD must work in its linear region to guarantee the accuracy of detection.

The nonlinearity of the BHD is reflected in two aspects including the PDs and the electrical amplifiers. Measuring the linear performance of the two parts separately cannot directly reflect the linear performance of the BHD, therefore we treat them as a whole. During the test, the pulsed light with 50 MHz and 50\% duty is generated by loading pulse signal output from the AWG on the AM modulator. The pulsed light is sent to only one PD of the BHD with the other blocked. We measure the output voltage of BHD by an oscilloscope and record the incident optical power. Meanwhile, we keep the optical power unchanged and measure the output voltage again with another PD illuminated. At the different optical power ranging from 2 uW to 57 uW, the output voltage of the two PDs is recorded and illustrated in Fig. 4 (a). The trans-impedance gain, which is equivalent to the slope of the fitted line, is calculated as 4.86 k. And the PDs are working in their linear region with only 2.9\% deviation\footnote{The deviation is calculated here as the standard deviation of each data point relative to the fitted line.} from the fitted line up to 57 uW \cite{chi2011balanced}.

\subsection{CMRR}

To quantify the subtraction capability of the BHD, we introduce the CMRR which is calculated by calculating the difference value between differential mode signal and common mode signal in the frequency domain. As demonstrated in Fig. 1 (b), the CMRR is obtained by measuring the output spectral power of the BHD at the pulsed light of 50 MHz in two cases: (a) only one PD is illuminated and another is blocked, (b) both PDs are illuminated. Since the CMRR has no relation to the input optical power \cite{chi2009high}, we set the input optical power as 41.8 uW to avoid the BHD saturation in case (a). In order to eliminate the common mode signal as much as possible, one needs to adjust VOA$_2$ and VOD finely to get a smaller residual signal according to the output of the BHD in case (b). The measured spectral power for both cases is displayed in Fig. 4 (b). The red curve represents the differential mode signal when only one PD is illuminated and the blue curve represents the common mode signal when both PDs are illuminated. The CMRR can be calculated based on the maximum difference of the fundamental harmonic spectral power. It is clear that the CMRR of our BHD reaches 57.9 dB which exceeds the CMRR in many reported detectors \cite{gray1998photodetector,2012Improvement,kumar2012versatile,chi2011balanced,duan2013300,mahler2017chip}.

\begin{figure}[t]
\centering
\includegraphics[width=35pc]{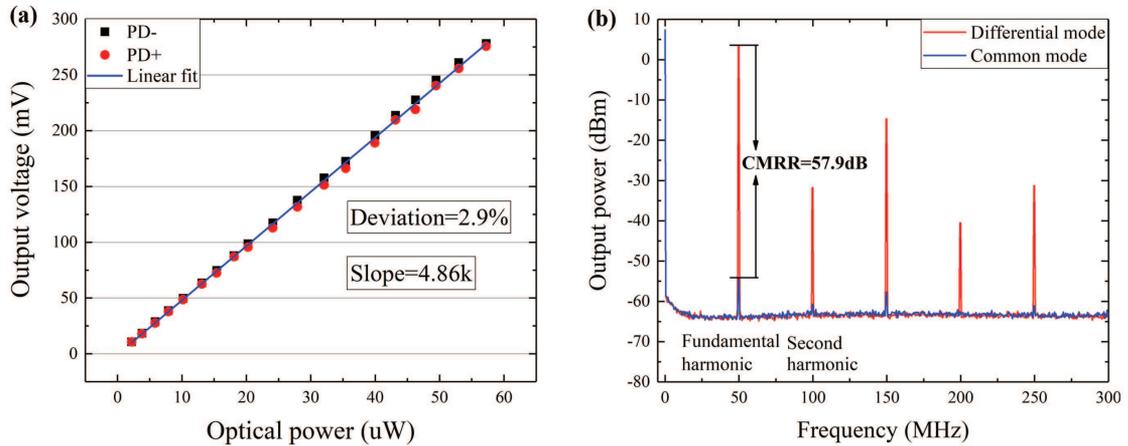}
\caption{(a) BHD output peak voltage as a function of the optical power. The "PD+" and "PD-" are reversely biased photodiodes that generate opposite direction of the current. The black dots are the output voltage of the BHD when only "PD-" is illuminated and the red dots are the output voltage of the BHD when only "PD+" is illuminated. The blue line represents the fitting line of the recorded voltage value. (b) Noise spectrum at an LO power of 41.8 uW. Resolution bandwidth: 100 kHz. The red curve is differential mode signal when only one PD is illuminated and the blue curve is the common mode signal when double PDs are illuminated and finely balanced. The peak located in 50 MHz is the fundamental harmonic while others peaks are higher harmonic.}
\label{fig_env5}\vspace*{-6pt}
\end{figure}

\subsection{Comparison between various BHDs}
We summarize a variety of BHDs used in CV-QKD or QRNG, and point out several key parameters, such as telecommunication wavelength, bandwidth, QCNR, and CMRR. A comparison of the specifications between our BHD and other BHDs reported in the literature is given in Table 1. In the parameters that characterize the BHD, the QCNR and bandwidth should primarily be considered. As described in Fig. 5, we show the sketch map of QCNR and bandwidth of the BHDs which are listed in Table 1. As shown, our BHD is no less impressive than its counterparts: it reflects a unique combination of the bandwidth and QCNR.

\begin{table}[!t]
\centering
\caption{A comparison between BHDs}
\label{tab1}
\begin{IEEEeqnarraybox}[\IEEEeqnarraystrutmode\IEEEeqnarraystrutsizeadd{2pt}{1pt}]{v/c/v/c/v/c/v/r/v/c/v/c/v/c/v/c/v/c/v/c/v/c/v}
\IEEEeqnarrayrulerow\\
& \mbox{BHD}
&&\mbox{\cite{hansen2001ultrasensitive}}
&&\mbox{\cite{jin2015balanced}}
&&\mbox{\cite{haderka2009fast}}
&&\mbox{\cite{cooper2013high}}
&&\mbox{\cite{kumar2012versatile}}
&&\mbox{\cite{chi2011balanced}}
&&\mbox{\cite{xu2012ultrastable}}
&&\mbox{\cite{mahler2017chip}}
&&\mbox{\cite{duan2013300}}
&&\mbox{Ours}&\\
\IEEEeqnarraydblrulerow\\
\IEEEeqnarrayseprow[3pt]\\
& \mbox{Wavelength (nm)} && \mbox{790} && \mbox{1064} && \mbox{800}&& \mbox{830} && \mbox{791} && \mbox{1550} && \mbox{577}&& \mbox{-}&& \mbox{1550}&& \mbox{1550}
&\IEEEeqnarraystrutsize{0pt}{0pt}\\
\IEEEeqnarrayseprow[3pt]\\
\IEEEeqnarrayrulerow\\
\IEEEeqnarrayseprow[3pt]\\
& \mbox{Bandwidth (MHz)} && \mbox{1}&& \mbox{2}&& \mbox{54}&& \mbox{80}&& \mbox{100}&& \mbox{104}&& \mbox{140}&& \mbox{150}&&\mbox{300}&&\mbox{1200}&\IEEEeqnarraystrutsize{0pt}{0pt}\\
\IEEEeqnarrayseprow[3pt]\\
\IEEEeqnarrayrulerow\\
\IEEEeqnarrayseprow[3pt]\\
& \mbox{QCNR (dB)} && \mbox{14}&& \mbox{37}&& \mbox{12} && \mbox{14.5}&& \mbox{13}&& \mbox{13}&& \mbox{10}&& \mbox{10}&& \mbox{14}&& \mbox{18.5}&\IEEEeqnarraystrutsize{0pt}{0pt}\\
\IEEEeqnarrayseprow[3pt]\\
\IEEEeqnarrayrulerow\\
\IEEEeqnarrayseprow[3pt]\\
& \mbox{CMRR} && \mbox{85}&&\mbox{75.2}&&\mbox{47.4} && \mbox{63}&& \mbox{52.4}&& \mbox{46}&& \mbox{55}&& \mbox{28}&& \mbox{54}&& \mbox{57.9} &\IEEEeqnarraystrutsize{0pt}{0pt}\\
\IEEEeqnarrayseprow[3pt]\\
\IEEEeqnarrayrulerow
\end{IEEEeqnarraybox}
\end{table}

\begin{figure}[t]
\centering
\includegraphics[width=36pc]{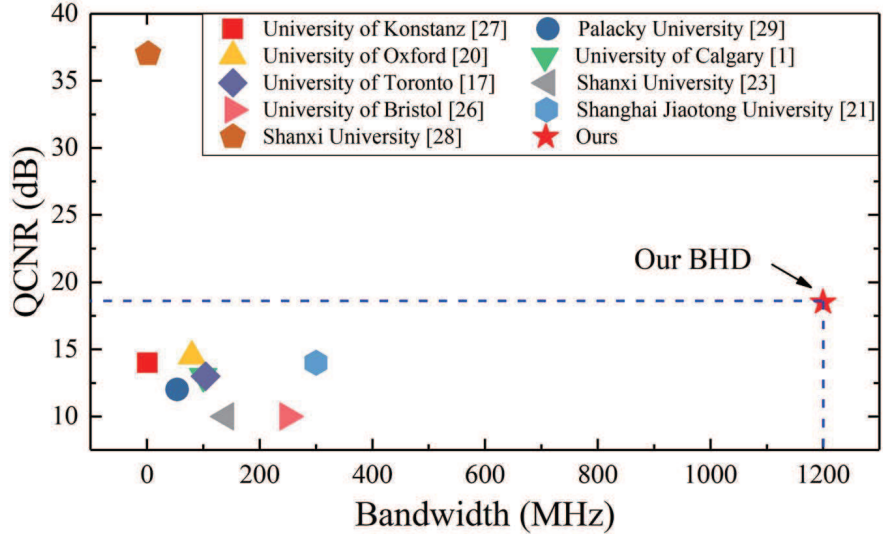}
\caption{Sketch map of quantum to classical noise ratio and bandwidth in the reported BHDs and our BHD. The arrow indicates our BHD.}
\label{fig_env8}\vspace*{-6pt}
\end{figure}

\section{Application}
We have characterized the parameters of the BHD whose QCNR is more than 10 dB and the bandwidth is as high as 1.2 GHz. Given this practical BHD, we further simulate the secret key rate in CV-QKD and analyze the random number generation rate in QRNG.

\subsection{BHD application in CV-QKD}
Under the protocol of Gaussian-modulated coherent-state CV-QKD \cite{braunstein2005quantum,weedbrook2012gaussian}, Alice encodes the key information (random bits with Gaussian distribution) by modulating field quadratures of weak coherent states. Experimentally, this is realized by modulating the intensity and the phase of each pulse. Over a distance, the BHD is used to measure the field quadratures on Bob's side. When the transmission is complete, Alice and Bob perform postprocessing to distill secure keys \cite{van2004reconciliation,bennett1995generalized,wang2018high}. This protocol has an advantage in achieving higher secure key per pulse than single photon protocol, while suffers from low repetition rate under the limitation of the bandwidth of BHD\footnote{The repetition rate should never exceed the bandwidth of the BHD.}. Owing to the requirement, most research institutions have developed their own BHDs displayed in table 1. Yet for all that, the repetition rate is still under the dome of hundreds of MHz while our BHD has potential to make a breakthrough for its bandwidth of 1.2 GHz.

To confirm the relationship between the parameters of our BHD and the maximal secret key rate, we simulate the secret key rate for a practical CV-QKD system when the repetition rate is 1.2 GHz. As drawn in Fig. 6, the dark red solid curve, dark blue dot-dashed curve and dark black dotted curve correspond to the data lengths of $N=10^8, N=10^9, N=10^{10}$, respectively. It can be seen that the secret key rate reaches 66.55 Mbps at a transmission distance of 10 km when considering a short data length $N=10^8$, which is much higher than the reported 10 Mbps at this distance \cite{inproceedings}. When the transmission distance is up to 45km, the secret key rate in our application is also as high as 2.878 Mbps compared with 301 kbps in \cite{dixon2015high}.

\begin{figure}[t]
\centering
\includegraphics[width=30pc]{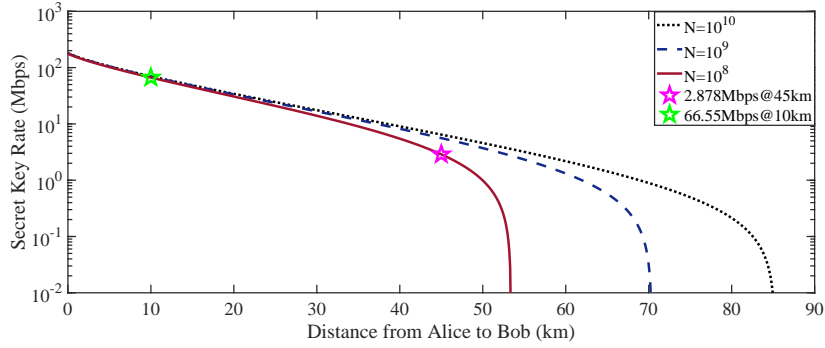}
\caption{Secure key rate as a function of the transmission distance based on our BHD under the finite-size effect. The simulation parameters include $\eta=0.612$, $\upsilon_{ele} =0.1 $ (in units of quantum noise), $V_A=2$, $\epsilon_A=0.043$ and $\beta=0.95$. The data lengths from left to right curves correspond to $N=10^8,N=10^9,N=10^{10}$ \cite{zhang2017continuous}.}
\label{fig_env9}\vspace*{-6pt}
\end{figure}

\subsection{BHD application in QRNG}
In the framework of measuring vacuum fluctuations of an electromagnetic field as the source of randomness \cite{gabriel2010generator,symul2011real,shen2010practical}, a QRNG contains entropy source, detection, sampling and randomness extraction. In our experiment, an actual BHD is applied to perform the detection of quadrature amplitude of the vacuum state. We are going to focus on analyzing the impact of BHD parameters on random number generation rate.

For a practical QRNG system, we assume that the measured noise variance $\sigma ^2_M$ is the sum of independent electronic noise variance $\sigma ^2_E$ and quantum noise variance $\sigma ^2_Q$ \cite{xu2012ultrastable}. The ideal value $\sigma ^2_Q$ is selected in the optical experiment when the measured variance $\sigma ^2_M$ is largest at the same time the devices are working in linear region \cite{xu2012ultrastable}.

We take the setup in Refs. \cite{zheng2018} as an example to evaluate the process of QRNG. As previously mentioned, the sampling rate, sampling accuracy, and reference voltage of ADC are 1 GHz, 12 bits, and 1.5 V respectively. We record the output voltage along with the LO power from 0 mW with a step size of 0.35 mW to 10.5 mW as raw data. Then we calculate the voltage variance of the raw data and normalize the voltage variance with the reference voltage of ADC. The trend is expressed in Fig. 7. It indicates that the voltage variance enhances linearly with the increase of the LO power ranging from 0 mW to 9.45mW. The maximum LO power of this linear region appears at 9.45 mW, where $\sigma ^2_M = 1386.59$, $\sigma ^2_E = 21.71$ and $\sigma ^2_Q = \sigma ^2_M - \sigma ^2_E = 1364.88$ ($\sigma ^2_E$, $\sigma ^2_Q$, $\sigma ^2_M$ are variance of raw data without normalization). To estimate extractable quantum randomness, we refer to the notion of min-entropy \cite{haw2015maximization}. Hence the extractable randomness of our measurement outcomes conditioned on classical noise can be described as min-entropy, which is
\begin{equation}
H_{min}(M|E) =log_2(2\pi\sigma ^2_Q)^{1/2}
\label{eq2}
\end{equation}
This yields a min-entropy of 6.53 bits per 12-bit sample. Our results reveal a potential random generation rate of 6.53 Gbps.

\begin{figure}[t]
\centering
\includegraphics[width=30pc]{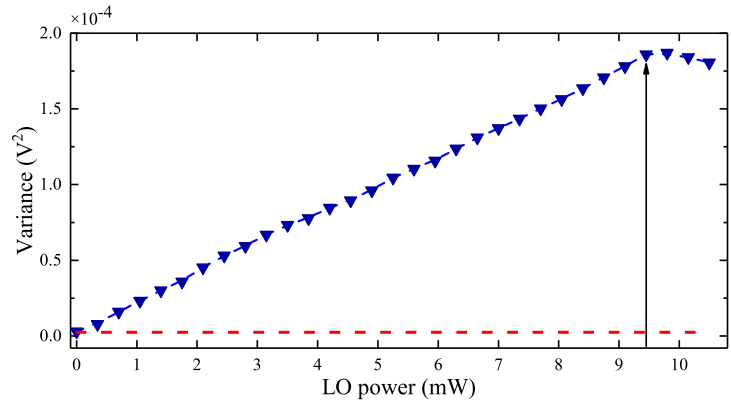}
\caption{Voltage variance as a function of LO power with ADC\&FPGA taking samples. In the region of 0 mW to 9.45 mW, it shows a linearity between the voltage variance and the LO power. When the LO power increases, the BHD appears to saturate, and the peak value of the voltage variance in linear region is obtained at 9.45 mW. The Y-axis is quantized with a 12-bit sampling rate and a 1.5 V reference voltage.}
\label{fig_env10}\vspace*{-6pt}
\end{figure}

\section{Conclusions}
In this paper, we have developed a 1.2 GHz balanced homodyne detector based on radio frequency and integrated circuit technology. This method is superior to the general trans-impedance chip in extending the bandwidth. The balanced homodyne detector also achieves a quantum to classical noise ratio of 18 dB at a local oscillation power of $\sim$8 mW and a high common mode rejection ratio of 57.9 dB. As a demonstration, we have pointed that the high performance balanced homodyne detector makes it possible to achieve secret key rate of 66.55 Mbps and 2.878 Mbps at a transmission distance of 10 km and 45 km respectively in continuous-variable quantum key distribution, and the random number generation rate of 6.53 Gbps in quantum random number generator.

\small
\bibliographystyle{IEEEtran}
\bibliography{Referance}

\end{document}